\documentclass[aps,prd,twocolumn,nofootinbib,preprintnumbers,superscriptaddress]{revtex4-1}
\usepackage[utf8]{inputenc}
\usepackage{amssymb,amsmath, amsfonts,mathrsfs,enumerate}
\usepackage{graphicx,rotate}
\usepackage{color, hyperref, dcolumn}
\usepackage{float}
\usepackage{soul}
\usepackage{slashed}
\usepackage{mathtools}
\usepackage{multirow, subfig}

\makeatletter
\let\Hy@linktoc\Hy@linktoc@page
\makeatother

\definecolor{ourcolor}{rgb}{0.7, 0.25, 0.05}

\long\def\rpl#1!!#2!!{\textcolor{red}{#1} \textcolor{blue}{#2}}

\let\bar=\overline

\def \order(#1){{\mathcal O} \left(#1 \right)}

\evensidemargin=\oddsidemargin
\allowdisplaybreaks

\begin{document}

\title{Multi-messenger Astronomy with Quenched Superradiance}

\author{Indra Kumar Banerjee}
\email{indrab@iiserbpr.ac.in}
\affiliation{Department of Physical Sciences, Indian Institute of Science Education and Research (IISER)\\ Berhampur, Ganjam, Odisha, 760003, India}
\author{Soumya Bonthu}
\email{bsoumya@okstate.edu}
\affiliation{Department of Physics, Oklahoma State University, Stillwater, OK, 74078, USA}
\author{Ujjal Kumar Dey}
\email{ujjal@iiserbpr.ac.in}
\affiliation{Department of Physical Sciences, Indian Institute of Science Education and Research (IISER)\\ Berhampur, Ganjam, Odisha, 760003, India}

\begin{abstract}
We propose a novel method to study the ultra-light bosons, where compact rotating objects undergo the phenomenon of \textit{quenched} superradiance to create gravitational waves and neutrino flux signals. The neutrino flux results from appropriate coupling between the ultra-light bosons and the neutrinos. We consider a heavy sterile neutrino generation from ultralight scalar, which later results in active neutrino flux through neutrino oscillations, whereas we consider active neutrino generation directly from the vector bosons. We study the intertwining of gravitational waves and neutrino flux signals produced from a single source and elaborate if and when the signals can be detected in existing and upcoming experiments in a direct manner.
\end{abstract}


\maketitle

\section{Introduction}
\label{sec:intro}
The current parlance of fundamental physics suffers from the lack of information regarding two of its most important components, i.e. neutrinos and dark matter. The unanswered questions regarding neutrinos are the origin and absolute value of its mass as well as its nature i.e., whether they are Dirac- or Majorana-type particles. Many experimental efforts are underway to probe neutrinos from various sources across a large range of energy varying all the way from sub-meV to a few thousands of PeV~\cite{SAGE:1999zrn, MiniBooNE:2012meu, SciBooNE:2011qyf, Borexino:2013zhu,DUNE:2020fgq, Hyper-Kamiokande:2018ofw, KamLAND:2014gul,  SNO:2011hxd, MicroBooNE:2016pwy, T2K:2011qtm, IceCube:2013low, IceCube:2013cdw, ICAL:2015stm, RNO-G:2023kag, NOvA:2007rmc, Kamiokande:1986xwg, K2K:2000kji, ANTARES:1999fhm, MINOS:1998kez, Anderson:1998zza, CHOOZ:1997cow, Super-Kamiokande:1998wen, IceCube-Gen2:2020qha, DayaBay:2007fgu, Frejus:1994brq, RENO:2018dro, Andres:1999hm, DeBonis:2014jlo, KM3NeT:2018wnd, GRAND:2018iaj}. 
Interestingly, neutrinos can be an excellent messenger due to its abundance, originated from astrophysical, terrestrial, cosmological and other sources~\cite{Waxman:1997ti, Waxman:1998yy, Super-Kamiokande:2002hei, Schramm:1980xv, Barr:1989ru, Bellini:2013wsa, Mirizzi:2015eza, Davis:1968cp, Super-Kamiokande:2016yck, SNO:2002tuh, Ciscar-Monsalvatje:2024tvm, Herrera:2024upj}, and its weakly interacting nature which allows almost unhindered propagation of cosmological length scales. Consequently neutrinos can retain most of the information regarding their sources and the environment it travels through. Thus a lot of progress has been made to probe neutrinos from extra-terrestrial sources and many other experiments are being planned and invested on to widen the range of neutrinos that we can observe. 
For dark matter, the scope is even wider as there is a myriad of theories where dark matter can be of particulate nature, or some exotic compact object and even a combination of both. For the particulate dark matter, the mass can vary from $\mathcal{O}(10^{-24})$ eV to a few hundreds TeV~\cite{Griest:1989wd, Ferreira:2020fam}. 
For extremely tiny masses, i.e., in the ultra-light domain ($10^{-24}\mathrm{~eV}- 1\mathrm{~eV}$) it is extremely difficult to probe them directly in laboratory experiments and therefore we have to rely on indirect means. Naturally, the ultra-light bosons (ULB) here can be categorized into scalars and vectors. Scalars in this mass range specifically intrigues the community as the QCD axions and various axion-like particles (ALP) reside in this domain. For example, in string theoretic scenarios compactification can give rise to many different ALPs and they form an `axiverse', some of which also appear in this mass domain~\cite{Arvanitaki:2009fg}. 
There can be scalars even lighter than $10^{-24}\mathrm{~eV}$ and while they can not play the role of dark matter, they are still viable candidates for dark energy~\cite{Peebles:2002gy}. Furthermore, ultra-light scalars (and vectors) can also be the mediator of a long range `fifth force'~\cite{Fischbach:1992nm}. Due to this versatility of the ultra-light scalars (ULS), it is very interesting to probe them to gain information about their properties, more precisely their interaction with the standard model (SM) sector. While direct detection of ULS is infeasible with the currently available technology, the indirect effects originating from astrophysical or cosmic events, such as cosmic birefringence~\cite{Carroll:1989vb,Carroll:1991zs,Harari:1992ea}, axion(ALP)-photon conversion around a compact object~\cite{Raffelt:1987im}, gravitational waves~\cite{Arvanitaki:2010sy}, and neutrino flux~\cite{Chen:2023vkq} generated from superradiance around a compact object, etc. can come to the rescue. 
In this study, we focus on the gravitational waves (GW) and neutrino flux originating from superradiance around a compact object. However, the generation of neutrino flux depends upon the coupling, if any, between the neutrinos and the ultra-light bosons~\cite{Chen:2023vkq}. It is worth mentioning here that a coupling of this type between ultra-light bosonic field and neutrinos exists in various models involving generation of neutrino mass~\cite{Gelmini:1980re}, grand unified theory~\cite{Georgi:1974sy,Pati:1974yy,Mohapatra:1974hk} etc.
In the ultra-light vector front there are motivations from low-energy limit of quantum gravity theories~\cite{Goodsell:2009xc}, dark photons~\cite{Fabbrichesi:2020wbt}, additional symmetries in the leptonic sector~\cite{Foot:1990mn, He:1991qd, Foot:1994vd}, etc. Such scenarios get constraints from dark matter direct detection experiments~\cite{XENON100:2014csq, XENON:2019gfn, XENON:2020rca}, light shining through a wall experiments~\cite{Ehret:2010mh}, astrophysical and cosmological observations~\cite{Garcia:2020qrp, Witte:2020rvb, Redondo:2013lna, An:2013yfc, Joshipura:2003jh, KumarPoddar:2020kdz, KumarPoddar:2019ceq}, etc. 
As mentioned previously, due to the very low mass of these bosons, direct detection is almost an insurmountable challenge. Black hole superradiance can be one of the viable option for some kind of indirect observation. Superradiance is a proposed phenomenon in which bosons are spontaneously created around a rotating compact object~\cite{Bekenstein:1973mi,Bekenstein:1998nt}. In this study, we consider rotating black holes (BH) to be the originator of such superradiance and we consider ultra-light bosons that can form a cloud around the rotating BH under certain conditions forming a bound system with discrete energy levels~\cite{Detweiler:1980uk}. GW can be produced from this in two-fold ways, namely, due to the transition of ULB from one energy level to another, and their annihilation to the gravitons~\cite{Arvanitaki:2010sy}.  
In this regard, it is worth mentioning that, GW, which was just a theoretical prediction less than a decade ago, has been observed for the first time by the LIGO collaboration in 2016 and since then almost a hundred GW events have been observed, which were transient in nature and astrophysical in origin~\cite{LIGOScientific:2016aoc}. Recently the pulsar timing arrays (PTA) have shown hints of GWs which are of stochastic nature~\cite{NANOGrav:2023gor,NANOGrav:2023hde,EPTA:2023fyk,EPTA:2023sfo,EPTA:2023xxk,Reardon:2023gzh,Zic:2023gta,Reardon:2023zen,Xu:2023wog}, i.e. they are GW background which might be astrophysical or cosmological in nature. 
A number of upcoming interferometer based GW detectors e.g., LISA~\cite{LISA:2017pwj}, Taiji~\cite{Ruan:2018tsw}, DECIGO~\cite{Kawamura:2011zz}, BBO~\cite{Phinney:2004bbo}, ET~\cite{Punturo:2010zz}, CE~\cite{Reitze:2019iox}, and in the ultra-high frequency domain detectors based on the principle of mechanical resonators  with optically-levitated sensors (LSD)~\cite{Arvanitaki:2012cn,Aggarwal:2020umq}, GW-electromagnetic wave conversion~\cite{Gertsenshtein:1962gw,Bahre:2013ywa,Albrecht:2020ntd}, dark matter radio (DMR) pathfinder based on resonant L-C circuits ~\cite{Chaudhuri:2014dla,Silva-Feaver:2016qhh}, interferometry~\cite{Holometer:2016qoh,Akutsu:2008qv}, gaussian beam (GB)~\cite{Li:2003tv,Li:2004df,Li:2006sx}, radio telescope~\cite{Fixsen:2009xn,Bowman:2018yin} etc.  will only flourish this burgeoning field of GW astronomy.  
Moreover, multi-messenger astronomy with already existing astronomical messengers like photons, cosmic rays, neutrinos, will greatly be augmented by the arrival of GW astronomy.
In this article, we aim to probe the ultra-light bosons by considering its coupling with the active (for vectors) and sterile (for scalars) neutrinos (if any) through multi-messenger astronomy. The main scheme is as follows. We consider the source of ULB as the spinning BH which can undergo superradiant instability that leads to the formation of a cloud of ULB around them~\cite{Bekenstein:1973mi,Bekenstein:1998nt}. This formation however depends on the product of the masses of the ULB and the BH. We briefly discuss the mechanism in Sec.~\ref{sec:bhsuprad}. Next we consider the constituent messengers, i.e. (a) the transient GW originating from the annihilation of the bosons into gravitons~\cite{Arvanitaki:2010sy}, and (b) the neutrino flux which may be generated from the bosonic cloud provided that there is appropriate coupling between the neutrinos and the ULB~\cite{Chen:2023vkq}. The main feature of this mechanism is that depending on the variables, there might be a phase during the superradiance where the rate of energy extracted by the bosonic cloud from the black hole is balanced by the energy extracted from the cloud by the neutrino flux. The presence of this phase makes the superradiance last long at the expense of quenching the superradiance process, hence the name \textit{quenched superradiance}. We then discuss the dependence of these two signals, i.e. signals which are generated from the bosonic cloud, on various parameters of the source BH e.g. luminosity distance ($d$), mass ($M_{\mathrm{BH}}$), spin ($a_*$), and the properties of the bosons e.g. mass ($m_{\phi}$ for scalars and $m_{A^{\prime}}$ for vectors), coupling with the neutrinos ($g_{\phi\nu}$ for scalars and $g_{A^{\prime}\nu}$). One of the novel features of the study is that two messengers arising from the same source are considered in combination to determine the properties of ULB which would otherwise be impossible to probe. Furthermore, it is to be noted here that though these two messengers from BH superradiance have separately been studied, our study is the first one to give a prescription on how to effectively use both of them in consonance to shed some light on the ULB.

This article is organized as follows. In Sec. \ref{sec:bhsuprad} we briefly explain the basics of black hole superradiance followed by Sec.~\ref{sec:qsrbh} where we discuss the quenched superradiance mechanism in details. In Sec. \ref{sec:results} we explain our finding with the help of a few benchmark cases and finally in Sec. \ref{sec:concl} we summarize and conclude.

\section{Review of Black Hole Superradiance}
\label{sec:bhsuprad}
Black holes, which are some of the very first solutions of the Einstein's field equations, can be characterised mainly by three properties, i.e. their mass, charge, and spin. All black holes, irrespective of their mass, charge, and spin has an event horizon from which nothing, even photons can escape. Furthermore, rotating black holes, which are the main interest of this work, has a region outside their event horizon, inside which all objects are bound to co-rotate with the black hole. This region is called the ergoshpere and it is of specific interest because negative energy object can exist inside this region. Using this theoretical property of the ergosphere, Penrose conducted a thought experiment, in which an object of finite initial energy can enter inside the ergoregion, where it can decay into two daughter particles, one with negative and another with positive energy. If the negative energy daughter particle fall into the black hole then the positive energy daughter particle can exit the ergoregion with energy higher than the parent particle. This process of energy extraction from a rotating black hole, is called the Penrose process and the field theoretical analogue of this process forms the foundation of black hole superradiance.
Black hole superradiance is a phenomenon through which particles can form a cloud around a black holes at the expense of the energy of the black hole. In this case, we are considering the formation of a cloud of bosonic particles around rotating black holes~\cite{Bekenstein:1973mi,Bekenstein:1998nt}. Essentially the boson field amplifies around the compact object through the extraction of energy and angular momentum of the black hole. In this study, we focus on ULB cloud forming around rotating black holes. It is worth mentioning here that the two quantities through which a rotating black hole can be characterized are the mass of the black holes $M_{\mathrm{BH}}$ and the dimensionless spin of the black hole $a_* = J/GM_{\mathrm{BH}}^2$ where $J$ is the angular momentum of the black hole. 
The boson cloud can be gravitationally bound to the rotating BH to form a bound system with discrete energy levels. This system is analogous to a hydrogen atom and hence it is termed as `gravitational atom'. The efficiency of the superradiance process, i.e. the creation of the ULB cloud, depends on the ratio of the gravitational radius ($r_g = GM_{\mathrm{BH}}$) of the BH and the Compton wavelength of the boson. This ratio is called gravitational fine structure constant $\alpha_g$ and can be expressed as a combination of the mass of the boson and the BH~\cite{Brito:2015oca},
\begin{align}
\alpha_g=\dfrac{GM_{\mathrm{BH}}m_{\phi/A^{\prime}}}{\hbar c},
\label{eq:alphag}
\end{align}
where $G$ is the Newton's gravitational constant, and $m_{\phi}$ ($m_{A^{\prime}}$) is the mass of the scalar (vector) boson. The superradiance is negligible if the $\alpha_g$ is too small or too large\footnote{Both for $\alpha_g\gg 1$ or $\alpha_g\ll 1$, the superradiance rate is exponentially suppressed, hence we do not consider it.} whereas it is the most efficient when $\alpha_g\lesssim\mathcal{O}(1)$~\cite{Brito:2015oca}. However, we take the conservative approach and throughout this article, we consider $\alpha_g$ to be $\mathcal{O}(0.1)$.  From that aspect, we can find the mass range for the black holes in question which will superradiate ultra-light bosons (within the mass range between $10^{-24}$ eV and $1$ eV) to be between $10^{14}M_{\odot}$ and $10^{-10}M_{\odot}$. Therefore, very small portion of this range can be astrophysical in nature while most of the range is within the primordial or supermassive domain. However, we remain agnostic regarding the nature of the BH in this study. Once these conditions are met, superradiance occurs due to the transfer of angular momentum from the BH to the states of a massive boson, which amplifies the field value of the scalar in question. The ground state of the ULB, in the Boyer-Lindquist coordinates, takes the form~\cite{Brito:2015oca},
\begin{widetext}
\begin{align}
\Phi(\vec{x},t)&=\Psi_0(t)R_{\alpha_g}(r)\sin\theta \cos(m_{\phi}t-\mathrm{\phi}),\mathrm{\qquad \qquad \qquad \qquad \qquad \qquad \qquad scalar,}\\
A^{\prime\mu}(\vec{x},t)&=\Psi_0(t)e^{-\alpha_g^2 r/r_g}(\alpha_g\sin\theta\sin(m_{A^{\prime}}t-\phi), \cos(m_{A^{\prime}}t), \sin(m_{A^{\prime}}t),0),\mathrm{~~~~~ vector},
\label{phifield}
\end{align}
\end{widetext}
where $\Psi_0$ is the peak field value and $R_{\alpha_g}(r)$ is the normalized radial solution. After the superradiance is triggered, the ULB will eventually form the cloud at the rate of superradiance $\Gamma_{\mathrm{SR}}$ which is approximately $\alpha_g^8 a_* m_{\phi}/24$ for scalars and $4\alpha_g^6 a_* m_{A^{\prime}}$ for vectors. The cloud mass is related to the peak field value as~\cite{Chen:2023vkq},
\begin{align}
M_c = 
\begin{cases}
    \dfrac{186\Psi_0^2}{\alpha_g^3 m_{\phi}}& \text{scalar,}\\
    \dfrac{\pi\Psi_0^2}{\alpha_g^3 m_{A^{\prime}}} & \text{vector.}
\end{cases}
\end{align}
It is worth mentioning here that we consider $M_c\sim 0.1M_{\mathrm{BH}}$ because if the cloud is heavier than this then the spin of BH will go lower than the threshold required for superradiance and it will come to a halt~\cite{Chen:2023vkq}. This threshold, i.e., the condition where superradiance is active can be expressed as\footnote{It is to be noted here that the gravitational atom has a structure similar to the hydrogen atom, i.e., the discrete energy states of the gravitational atom can be expressed with the quantum numbers $(n,l,m)$ which are the overtone, orbital, and azimuthal quantum numbers respectively. It is to be noted here that the principal quantum $\nu=n+l+1$. In this work, for the sake of simplicity, we only consider the most dominant unstable state which can be expressed as $n=0,~l=m=1$.},
\begin{align}
\alpha_g < \dfrac{a_*}{2\left(1+\sqrt{1-a_*^2}\right)}.
\end{align}
In this regard, one can easily express the peak value of the field value of the boson when cloud mass is $10\%$ of that of the black hole mass as,
\begin{align}
\Psi_{10\%} = 
\begin{cases}
    1.1\times 10^{25}\left(\dfrac{\alpha_g}{0.2}\right)^2\mathrm{~eV}& \text{scalar,}\\
    8.7\times 10^{25}\left(\dfrac{\alpha_g}{0.2}\right)^2\mathrm{~eV} & \text{vector.}
\end{cases}
\end{align}
It is worth mentioning that here we only discuss isolated black holes and not binaries. Superradiance around compact binaries can also have very interesting phenomenological implications as shown in Refs.~\cite{Zhang:2019eid, Cao:2023fyv, Tomaselli:2024dbw}.
\section{Quenched Superradiance of Black Hole}
\label{sec:qsrbh}
In the previous section we briefly discussed the vanilla scenario of black hole superradiance, i.e., where the ULBs are only gravitationally bound to the black hole and hence the boson cloud takes away energy and angular momentum from the black hole till the black holes spin is less than the critical value after which the superradiance stops. However, there could be other non-trivial scenarios where the bosons have self interactions and/or interactions with other SM and beyond standard model particles which can alter the superradiance mechanism. If the superradiacne is modified such that for some time, the energy extracted by the boson cloud from the black hole is the same as the energy extracted from the boson cloud by some other mechanism, then the superrdiance mechanism runs for much longer as compared to the vanilla scenario. We term this kind of scenario the quenched superradiance of black holes. Such quenching may significantly modify the evolution of the BH and cloud.
In this study, we consider a scenario where the energy extracted by the bosonic cloud from the BH is taken away from the cloud by the fermions, i.e. the fermion production rate and the superradiance rate becomes equal to each other at some point. In this regard, it is to be noted that the creation of fermions from the bosonic cloud depends on the nature of the bosons, i.e., if the cloud is made of scalars, then the fermions can be created through parametric excitation from the oscillating scalar background in the cloud, whereas, if the cloud is made of vector bosons, then the fermions are produced through Schwinger pair production. Here we only show the expressions for the rate of fermion production from both the mechanisms and we encourage the interested reader to refer to Ref.~\cite{Chen:2023vkq} for further details on these mechanisms.
For the scalar case, if a fermion $f$ has a coupling with the scalar boson of the form $g_s\phi \bar{f}f$, then the production rate of fermions from the parametric excitation can be expressed as,
\begin{align}
\Gamma_{s} = \dfrac{g_s^2\phi_0^2 m_{\phi}^2}{48\pi^3}\sqrt{\dfrac{m_{\phi}}{g_s\phi_0}}.
\end{align}
For the case of vector bosons, if the coupling between the boson and the fermion is of the form $g_V A^{\prime\mu}\bar{f}\gamma_{\mu} f$ the rate of fermion production per unit time and unit volume through Schwinger pair production can be expressed as,
\begin{align}
\Gamma_{V} = \dfrac{g_V^2 E_{A^{\prime 2}}}{48\pi},
\end{align}
where $E_{A^{\prime}}$ is the `electric' field strength associated with the vector boson $A^{\prime}$ and can be expressed as $E_{A^{\prime}} \approx m_{A^{\prime}}|\vec{A^{\prime}}|$. It is to be noted here that the Pauli exclusion principle does not affect the fermion production since after the creation of a fermion it is accelerated to very high energies and far away from its original position which does not interfere in the creation of a subsequent fermion. 
It was shown in Ref.~\cite{Chen:2023vkq} that these fermions can have very high energies which can be expressed as,
\begin{align}
E_f = 
\begin{cases}
    0.27g_s\Psi_0,& \text{scalar},\\
    0.35g_V\Psi_0 & \text{vector},
\end{cases}
\label{ferme}
\end{align}
whereas the differential flux of these fermions to an observer at distance $d$ can be expressed as,
\begin{widetext}
\begin{align}
\Phi_f = 
\begin{cases}
   1.2\times 10^{-17}\left(\frac{\Psi_0}{4.8\times 10^7\mathrm{~GeV}}\right)^{1/2}\left(\frac{N_f}{3}\right)\left(\frac{10^{-12}\mathrm{~eV}}{m_{\phi}}\right)^{1/2}\left(\frac{0.3}{\alpha_g}\right)^3\left(\frac{g_s}{10^{-8}}\right)^{1/2}\left(\frac{5\mathrm{~kpc}}{d}\right)^2\mathrm{~cm}^{-2}\mathrm{s}^{-1}\mathrm{eV}^{-1}& \text{scalar},\\
    1.3\times 10^{-6}\left(\frac{\Psi_0}{5.7\times 10^{14}\mathrm{~GeV}}\right)\left(\frac{N_f}{1}\right)\left(\frac{10^{-12}\mathrm{~eV}}{m_{A^{\prime}}}\right)\left(\frac{0.3}{\alpha_g}\right)^3\left(\frac{g_V}{10^{-12}}\right)\left(\frac{5\mathrm{~kpc}}{d}\right)^2\mathrm{~cm}^{-2}\mathrm{s}^{-1}\mathrm{eV}^{-1} & \text{vector},
\end{cases}
\label{fermflux}
\end{align}
\end{widetext}
where $N_f$ is the number of fermion species which is being produced from the bosons cloud. 

Finally, it is evident that in order for the quenched superradiance mechanism to work, i.e., for the rate of superradiance to be equal to the rate of fermion production, the field value has to have a `critical' value depending on the coupling between the boson and the fermion, the boson mass and the gravitational fine structure constant. The authors in Ref.~\cite{Chen:2023vkq} finds this critical value of the peak value of the boson field to be,
\begin{widetext}
\begin{align}
\Psi_0^c = 
\begin{cases}
    4.8\times 10^{16}\left(\frac{3}{N_f}\right)^2\left(\frac{m_{\phi}}{10^{-12}\mathrm{~eV}}\right)\left(\frac{\alpha_g}{0.3}\right)^{16}\left(\frac{10^{-7}}{g_s}\right)^5\left(\frac{a_*}{0.9}\right)^2\mathrm{~eV} & \text{scalar},\\
    5.7\times 10^{23}\left(\frac{1}{N_f}\right)\left(\frac{m_{A^{\prime}}}{10^{-12}\mathrm{~eV}}\right)\left(\frac{\alpha_g}{0.3}\right)^{6}\left(\frac{10^{-12}}{g_s}\right)^3\left(\frac{a_*}{0.9}\right)\mathrm{~eV} & \text{vector}.
\end{cases}
\end{align}
\end{widetext}
In the remaining part of this work, we discuss the scenarios where this is possible, under which conditions this happens and how the superradiance mechanism progresses once this condition is achieved.
\subsection{Conditions of Quenched Superradiance}
\label{subsec:condqsr}
There are two main conditions for the quenched superradiance to occur, i.e., (i) the field value when the superradiance gets quenched must be much less than the field value $\Psi_{10\%}$, (ii) the field value should be high enough for the production of the fermions. 
For the first condition, we consider the limit $\Psi_{10\%}$ because if the cloud mass reaches $10\%$ of that of the BH, the spin of the BH reduces significantly and the superradiance process comes to a halt. The first condition can be expressed as $M_c \ll 0.1 M_{\mathrm{BH}},~\Psi_c \ll \Psi_{10\%}$.

\begin{figure*}
	\centering
	\includegraphics[scale=0.42]{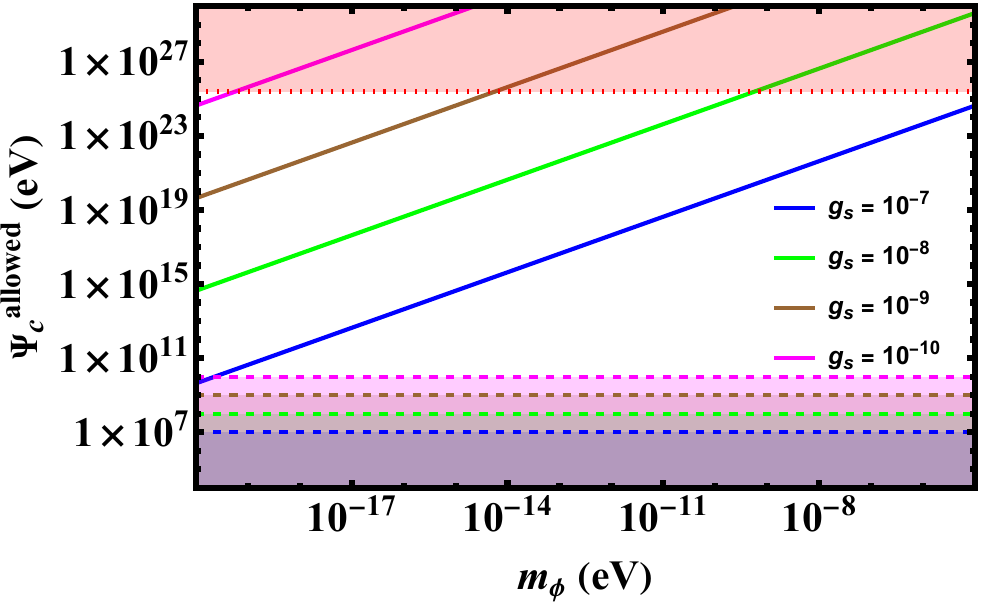}~~~~~~
	\includegraphics[scale=0.42]{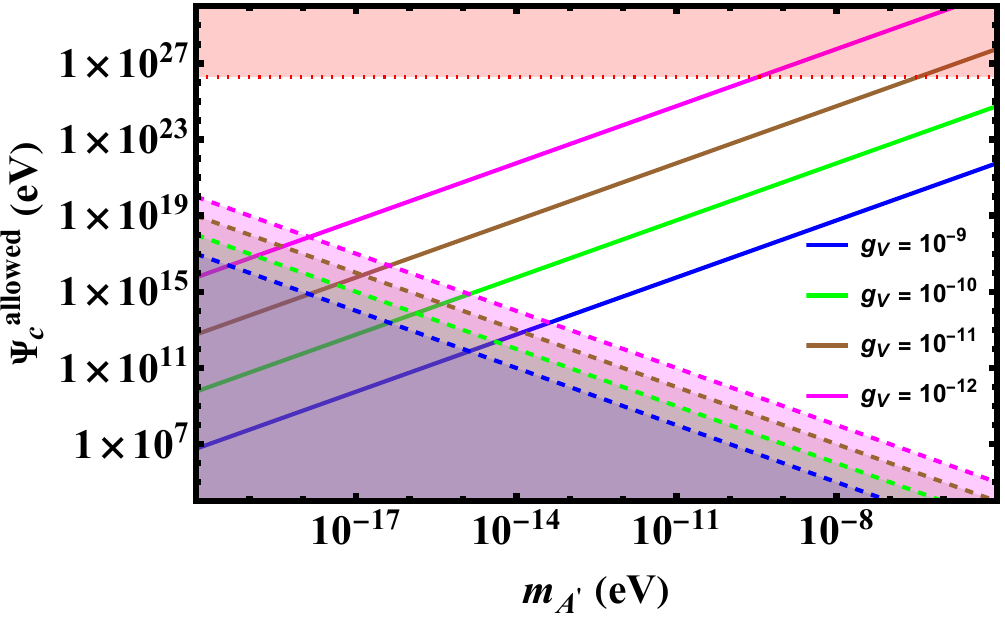}
	\caption{In the left (right) panel we show the allowed value of critical field value for which the quenched superradiance occur for a scalar (vector) boson case with respect to varying scalar (vector) mass and coupling between the scalar (vector) and the sterile (lightest active) neutrino. The upper bound here is for the first condition and the lower bound here is for the second condition, i.e., the condition for the parametric production.}
	\label{condsca}
\end{figure*}

The second condition differs for scalar and vectors. For the scalar cloud, the fermions are produced by parametric excitation, which requires $g_s\Psi_c\gg m_{f}$ where $m_f$ is the mass of the fermion that is being produced. 
In this study we concern ourselves with the production of detectable neutrino flux. Hence, we consider the fermions to be neutrinos. However, for the scalar case, we can not consider coupling between the scalars and active neutrinos in order to ensure efficient transaction of energy. This is due to the fact that when considering active neutrino states, due to the varying background scalar field, the active neutrinos may realise very high effective mass. As a result, immediately after production, these neutrinos may decay to neutral pions which have a mass of $\sim 140$ MeV, if the peak field value surpasses a few hundred MeV. Therefore, for situations with higher peak field value, the energy that is extracted form the scalar cloud in the form of active neutrinos, is not enough to balance the superradiance, making the quench impossible. Hence for the scalar case, we consider the fermion to be a sterile neutrino with mass $m_{\nu_s} = 1$ eV. These sterile neutrinos while propagating can oscillate into detectable active one.
For the vector boson case, the fermion generation mechanism is Schwinger pair production, for which the condition reads $m_{A^{\prime}} g_V\Psi_c\gg m_f^2$, where $m_{A^{\prime}}$ is the mass of the vector boson. Unlike the scalar case, the problem of efficient energy transaction does not occur in case of a coupling between active neutrinos and the vectors. Therefore, for the vector case, we take $m_{\nu_1} = 10^{-5}$ eV, i.e. we consider the fermion to be the lightest active neutrino.
In the left (right) panel of Fig.~\ref{condsca} we show the allowed parameter space for the scalar (vector) case based on the two above-mentioned conditions with respect to the mass of the boson and the coupling between bosons and neutrinos.
It can be seen from the left panel Fig.~\ref{condsca} that for very light scalars ($\mathcal{O}(10^{-20})$ eV) weaker coupling facilitates quenched superradiance whereas for less light scalars the situation is opposite. Furthermore, as can seen from the figure, in the scalar mass range that we consider, i.e., $m_{\phi}\in [10^{-20},10^{-5}]$ eV, $g_s =10^{-7}$ leads to the most amount of allowed region whereas weaker coupling violates the upper bounds. This is due to the fact that weaker coupling leads to lower production rate of the neutrinos and hence the balance of the the energy extraction is difficult to achieve. In this case, both lower and upper bounds are independent of the scalar mass as can be understood from the conditions discussed above.
For the case of vectors, the main difference is that the lower bound depends on the vector mass, as can be seen from the right panel Fig.~\ref{condsca}, more precisely, for lower vector mass, the lower bound becomes more stringent, leaving less room for the quenched superradiance to take place. For the vector case, weaker coupling cases violate the upper bound for heavier vectors whereas the stronger coupling cases violate the lower bound in the parameter space that we concern ourselves with. In the subsequent part of this work, we only remain within the unconstrained region of the parameter space, i.e., the region which is not bounded by the shaded regions in Fig.~\ref{condsca}, while performing further analysis.
\subsection{Time-lines in Quenched Superradiance}
\label{subsec:tlqsr}
In a quenched superradiance case depending on the situation, there can be three ways in which energy can flow inward or outward from the boson cloud, i.e. (i) the superradiance energy extraction by the cloud from the black hole, (ii) energy extracted from the cloud by the fermions, and (iii) the energy lost by the cloud in the form of gravitational waves. All these mechanisms of energy extraction change the mass of the boson cloud. The first leads to the gain of cloud mass whereas the second and the third one leads to the loss of the same.

In a quenched superradiance scenario, the evolution of the mass of the bosonic cloud is governed as,
\begin{align}
\dfrac{dM_c}{dt} = \Gamma_{\mathrm{SR}}M_c - 2E_f\int\Gamma_{s/V}d^3 x - \dfrac{dE_{\mathrm{GW}}}{dt},
\label{eq:evoleq}
\end{align}
where the first term in the RHS signifies the superradiant growth, the second term signifies rate of fermion production and the third term signifies the energy taken away by the gravitational waves.

The entire quenched superradiance regime can be divided into three main phases, i.e.,
\begin{itemize}
\item \textbf{The growth phase:} In this phase the fermion production rate is much less than the rate of superradiance, hence the evolution of the cloud depends only on the superradiance. This can be expressed as,
\begin{align}
\dfrac{dM_c}{dt} = \Gamma_{\mathrm{SR}}M_c,
\end{align}  
where $\Gamma_{\mathrm{SR}}$ takes the form $\alpha_g^8 a_{*} m_{\phi}/24$ for the scalar case and $4\alpha_g^6 a_* m_{A^{\prime}}$ for the vector case as mentioned before. This goes on till the field value of the bosonic cloud reaches the value of the critical field $\Psi_c$. The growth time, i.e. the time for which this process goes on before getting quenched can be expressed as,
\begin{align}
\tau_{\mathrm{growth}} = \dfrac{1}{\Gamma_{\mathrm{SR}}}\ln\left(\dfrac{C_{\phi/A^{\prime}}\Psi_c^2}{\alpha_g^3m_{\phi/A^{\prime}}}\right),
\end{align}
where $C_{\phi} = 186$ and $C_{A^{\prime}} = \pi$ as mentioned before. It is to be noted that since at the end of the growth phase, the mass of the cloud is many orders of magnitude less than the mass of the BH (which is one of the conditions for as mentioned in Sec.~\ref{subsec:condqsr}), we assume that the spin of the BH remains unchanged in this phase. 

\item \textbf{The balanced phase:} This is the phase which causes the quenched superradiance. In this phase the field value of the bosonic field takes such a form that the rate of superradiance and the rate of fermion production is the same. The rate of fermion production is given by $2E_f\int\Gamma_{s/V}d^3x$. In this case the cloud mass remains constant at $M_c = (C_{\phi/A^{\prime}}\Psi_c^2)/(\alpha^3_gm_{\phi/A^{\prime}})$. However. the spin of the BH evolves as,
\begin{align}
\dfrac{da_*}{dt} = -\dfrac{m_{\phi/A^{\prime}}}{\alpha_g^2}\Gamma_{\mathrm{SR}}M_c.
\end{align}
Now, since the spin of the BH reduces at a constant rate, and superradiance only works as long as the condition $\alpha_g< a_*/2(1+\sqrt{1-a_*^2})$ as mentioned before. Therefore, for a fixed value of $\alpha_g$, as soon as the spin realises a value which violates the above condition, the superradiance stops, and as a result, the balance breaks. The duration for which this phase lasts is denoted by $\tau_{\mathrm{balanced}}$ is the subsequent parts of the text.
\item \textbf{The depletion phase:} After the superradiance stops, the bosonic clouds do not have any more input in energy. However, since they still have a rather high value of the field (can be calculated my inverting the relation $M_c = (C_{\phi/A^{\prime}}\Psi_0^2)/(\alpha_g^3m_{\phi/A^{\prime}})$, where $\Psi_0$ is the peak field value.). Therefore, the mass of the bosonic cloud still reduces as,
\begin{align}
\dfrac{dM_c}{dt} = -2E_f\int\Gamma_{s/V}d^3x.
\end{align}
Now this process goes on till the production condition (mentioned in the previous subsection) is violated.
\end{itemize}
It is to be noted here these three phases are the ones with the most phenomenological impact. However, after the completion of these three phases, the rest of the cloud is slowly depleted in the form of gravitational waves.\footnote{The amplitude of these gravitational waves is far weaker than the current or proposed capabilities of the experiments, and hence we do not discuss these.} It can be understood from Eq.~\eqref{eq:evoleq}. i.e., at the end of these three phases, the equation takes the form $dM_c/dt=-dE_{\mathrm{GW}}/dt$. However, since the the energy extraction in the form of GW is very low with respect to time, we do not consider that phase.

\begin{figure*}
\centering
\includegraphics[scale=0.5]{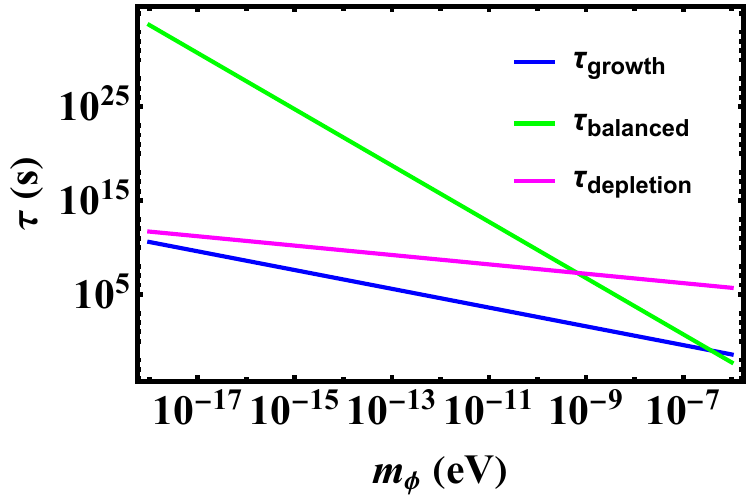}~~~~~~
\includegraphics[scale=0.5]{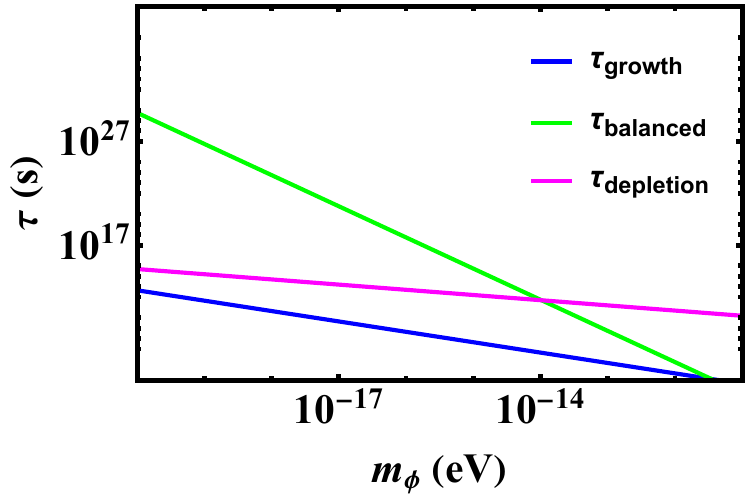}
\caption{The time-lines for the scalar case. In the left (right) panel, we have considered $g_s =10^{-7}~(10^{-8})$. Furthermore, for both the cases, we have taken $\alpha_g = 0.3$ and initial value of $a_* = 0.9$.}
\label{tlsca}
\end{figure*}

\begin{figure*}
\centering
\includegraphics[scale=0.44]{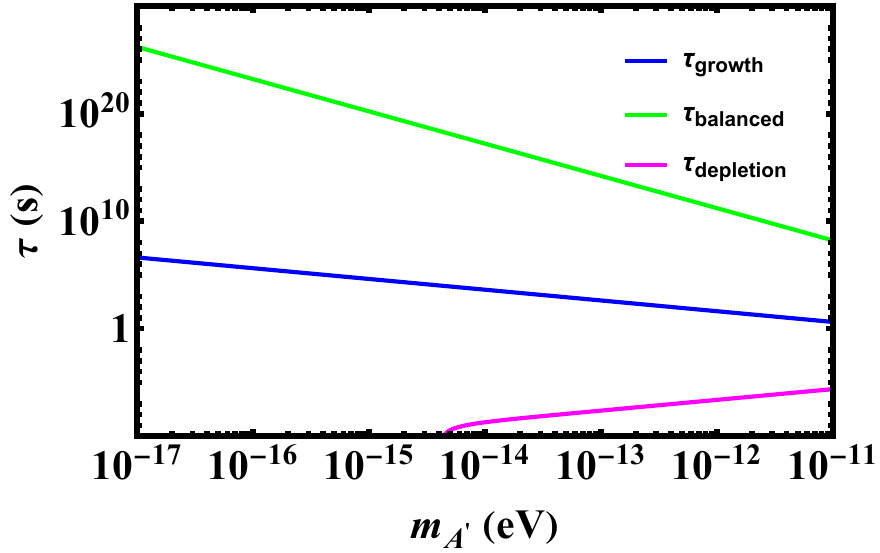}~~~~~~
\includegraphics[scale=0.446]{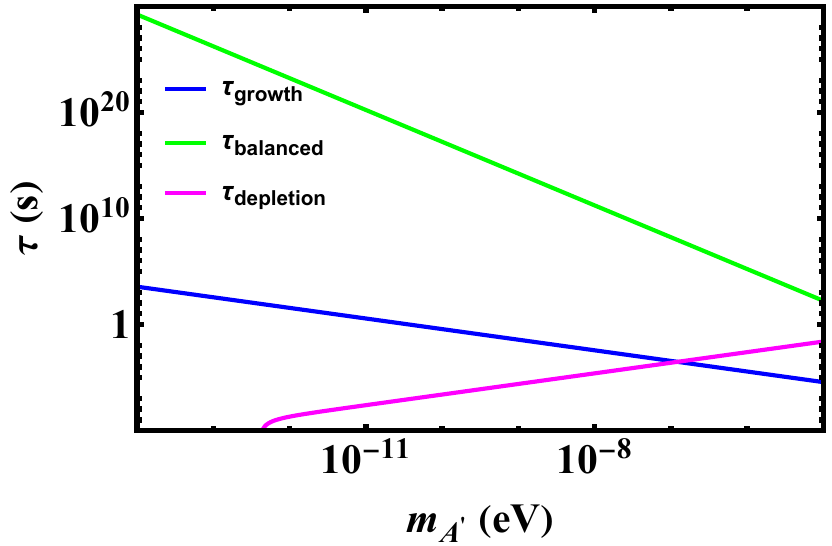}
\caption{The time-lines for the vector case. In the left (right) panel, we have considered $g_V =10^{-12}~(10^{-10})$. Furthermore, for both the cases, we have taken $\alpha_g = 0.3$ and initial value of $a_* = 0.9$.}
\label{tlvec}
\end{figure*}

\begin{figure*}
\centering
\includegraphics[scale=0.55]{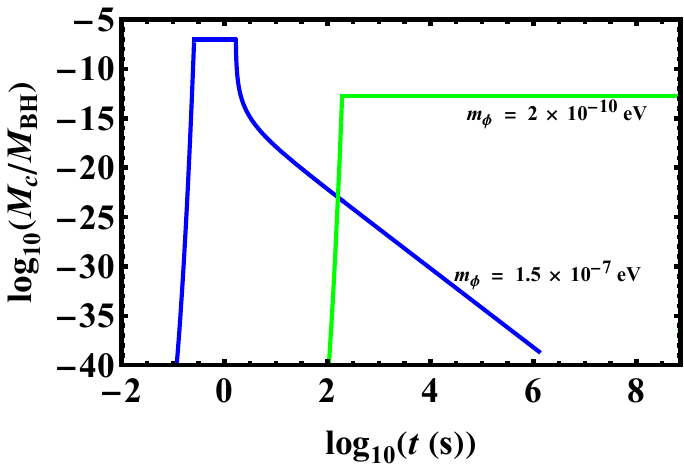}~~~~~~~~
\includegraphics[scale=0.5]{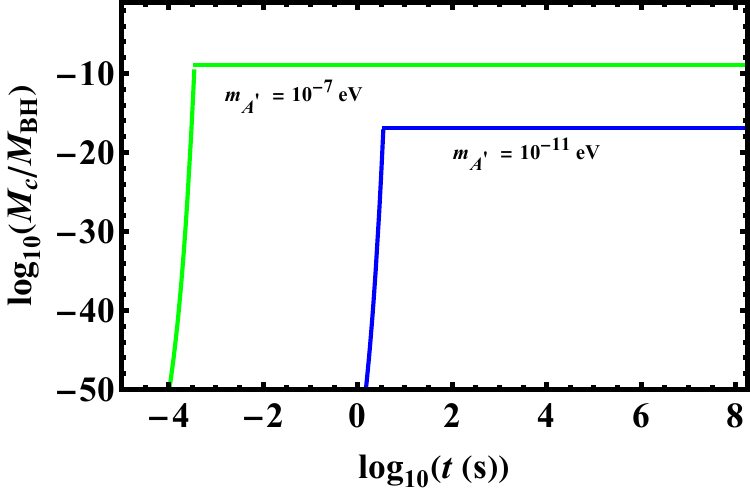}
\caption{(Left) Evolution of the scalar cloud with $m_{\phi}=2\times 10^{-10}$ eV (green) and $m_{\phi}=1.5\times 10^{-7}$ eV (blue). In both cases, $g_s=10^{-7}$, $\alpha_g = 0.3$ and initial $a_*=0.9$. (Right) Evolution of the vector cloud with $m_{A^{\prime}}=10^{-7}$ eV (green) and $m_{A^{\prime}}= 10^{-11}$ eV (blue). In both cases, $g_V=10^{-10}$, $\alpha_g = 0.3$ and initial $a_*=0.9$.}
\label{mcscavec}
\end{figure*}

In Figs. \ref{tlsca} and \ref{tlvec} we show the examples of such time-lines for the scalar and vector background respectively.
In the left (right) panel of Fig.~\ref{tlsca} we have shown the different phases mentioned above with $g_s = 10^{-7}~(10^{-8})$ with varying scalar mass. It can be seen that for the fixed value of scalar mass, $\tau_{\mathrm{growth}}$ and $\tau_{\mathrm{depletion}}$ are lower whereas $\tau_{\mathrm{balanced}}$ is higher for stronger couplings. This is due to the fact that for stronger coupling, the critical filed value is achieved faster and the balanced phase lasts longer. In Fig.~\ref{tlvec} in the left (right) panel we have shown the duration of the different phases of the quenched superradiance mechanism with the vector mass for coupling $g_V = 10^{-12}~(10^{-10})$. In this case, similar to the scalar case, we see that for stronger coupling the balanced phase lasts longer while the growth phase is shorter. However, unlike the scalar case, in the vector case the depletion phase is longer for the stronger coupling as for weaker coupling the particle production condition is more stringent. Furthermore, in Fig.~\ref{tlsca} we can see that there are regions in the parameter space for which the depletion phase lasts much longer than the balanced phase. In order to illustrate the comparison between these different phases, we show a few particular cases of evolution of scalar (vector) cloud in the left (right) panel of Fig.~\ref{mcscavec}.
In the left panel of Fig,~\ref{mcscavec} we have shown the evolution of a scalar cloud with scalar mass of $1.5\times 10^{-7}$ eV and $2\times 10^{-10}$ eV denoted by blue and green solid curves respectively. For both these cases, we have considered $g_s=10^{-7}$, $\alpha_g = 0.3$, and initial value of the black hole spin $a_* = 0.9$. One of the most important features which is to be noted in this figure is that in both the cases, the peak value of the cloud is much less than $10\%$ of the mass of the black hole, which is one of the conditions for the quenched superradiance. Furthermore, in the blue solid line one can see that though the duration of the growth phase is much smaller than the balanced phase, the duration of the depletion phase is even longer, which is not the same of the green solid line. This is aligned with the fact that for heavier scalars $\tau_{\mathrm{balanced}}< \tau_{\mathrm{depletion}}$ as shown in Fig.~\ref{tlsca}.

On the other hand we show the same thing for the vector clpud in the right panel of Fig.~\ref{mcscavec} with $g_V = 10^{-10}$, $\alpha_g=0.3$ and initial spin of the black hole $a_* = 0.9$ for vector masses of $10^{-7}$ eV and $10^{-11}$ eV denoted by the green and blue solid curves respectively. It is to be noted here that in this case, both curves correspond to a prolonged balanced phase as opposed to the heavier scalar case mentioned before.

%

\subsection{Signals from Quenched Superradiance}
\label{subsec:sigqsr}
In the previous subsections, we have discussed the conditions and  timelines of the quenched superradiance mechanism. Now we focus on the avenues to probe this scenario. In the quenched superradiance paradigm, the signals that we consider are, the neutrino flux and the GW strain. Furthermore, we consider these signals from the balanced phase of the quenched superradiance mechanism. This is due to the fact that both these signals increase with increasing mass of the bosonic cloud and during the balanced phase, mass of the cloud is the highest. Required details regarding the signals are explained in the subsequent text.
\begin{enumerate}
\item \textbf{Ultra high energy neutrino:} As mentioned before, in this work, we are considering a sterile neutrino $\nu_s$ of mass $1$ eV as the fermion for the scalar superradiance case where the coupling between the scalar and the neutrino can be expressed as $g_{\phi\nu}\phi\overline{\nu_s}\nu_s$. Therefore, one can expect flux of sterile neutrinos from a scalar cloud depending on the strength of the coupling, gravitational fine structure constant and the spin of the black hole. Although we can not directly detect sterile neutrino flux, we further assume that the sterile neutrinos mix with the active neutrinos and a result during their propagation towards the earth, the a small part of the sterile neutrino flux convert into active neutrinos which is detectable.
On the other hand for the vector cloud, we are considering a coupling between the vector and the lightest active neutrino state $\nu_1$, with mass $10^{-5}$ eV, of the form $g_{A^{\prime}\nu} A^{\prime\mu}\overline{\nu_1} \gamma_{\mu} \nu_1$. This lightest active neutrino mass flux can in turn be detected directly on earth.
All the required details regarding the flux and the energy of these neutrinos are already shown in Eqs.~\eqref{ferme} and \eqref{fermflux}.

\item \textbf{Transient GW signals:}
The GW that we consider in this study is created by the annihilation of the bosons into gravitons which can later be detected by the detector in the form of gravitational waves. 
\end{enumerate}
The GW frequency and RMS strain can be expressed as~\cite{Brito:2015oca},
\begin{align}
f_{\mathrm{GW}} = \dfrac{5000 ~m_{\phi/A^{\prime}}}{10^{-11}\mathrm{~eV}} \mathrm{~Hz},
~~
h_{\mathrm{rms}} = \sqrt{\dfrac{\frac{dE_{\mathrm{GW}}}{dt}}{5\pi^2 f_{\mathrm{GW}}^2 d^2}},
\end{align}
where,
\begin{align}
\dfrac{dE_{\mathrm{GW}}}{dt} = A_{\phi/A^{\prime}}\left(\dfrac{M_c}{M_{\mathrm{BH}}}\right)^2\alpha_g^{4l+10}.
\end{align}

Here both $A_{\phi}$ and $A_{A^{\prime}}$ is $\mathcal{O}(10^{-3})$~\cite{Yoshino:2013ofa,Baryakhtar:2017ngi}.
In this study, we consider the GW signal from the balanced phase, and therefore $h_{\mathrm{rms}}$ takes the form,
\begin{widetext}
\begin{align}
h_{\mathrm{rms}} = 
\begin{cases}
5.37\times 10^{-27}\left(\dfrac{m_{\phi}}{10^{-12}\mathrm{~eV}}\right)\left(\dfrac{\alpha_g}{0.3}\right)^{33}\left(\dfrac{\mathrm{kpc}}{d}\right)\left(\dfrac{10^{-8}}{g_{\phi\nu}}\right)^{10}\left(\dfrac{a_*}{0.9}\right)^4 & \text{scalar}\\
 1.95\times 10^{-32}\left(\dfrac{m_{A^{\prime}}}{10^{-12}\mathrm{~eV}}\right)\left(\dfrac{\alpha_g}{0.3}\right)^{14}\left(\dfrac{\mathrm{kpc}}{d}\right)\left(\dfrac{10^{-12}}{g_{A^{\prime}\nu}}\right)^{6}\left(\dfrac{a_*}{0.9}\right)^2 & \text{vector}.
\end{cases}
\end{align}
\end{widetext}
Furthermore, the characteristic strain can be expressed as,
\begin{align}
h_c = \sqrt{N_{\mathrm{cycles}}}h_{\mathrm{rms}},
\end{align}
where $N_{\mathrm{cycles}} = \mathrm{min}[fT_{\mathrm{obs}},f_s\tau_{\mathrm{GW}}]$, $f$ and $f_s$ are the GW frequency in the detector and the source frames respectively with the relation $f=f_s/(1+z)$; $T_{\mathrm{obs}}$ is the observation time of the relevant GW detector, and $\tau_{\mathrm{GW}}$ is the duration of the GW signal. In this study, we consider $\tau_{\mathrm{GW}}=\tau_{\mathrm{balanced}}$. 
\section{Results}
\label{sec:results}
We now discuss the two signals that can arise from a black hole going through quenched superradiance. At first we show on which part of the parameter space both the signals can be in the detectable range followed by a few benchmark cases.
\subsection{Combination of the Two Messengers}
\label{combi}
\begin{figure*}
\centering
\includegraphics[scale=0.57]{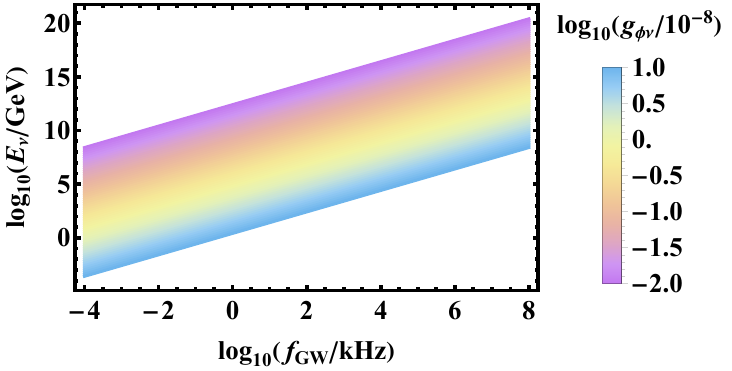}~~~
\includegraphics[scale=0.57]{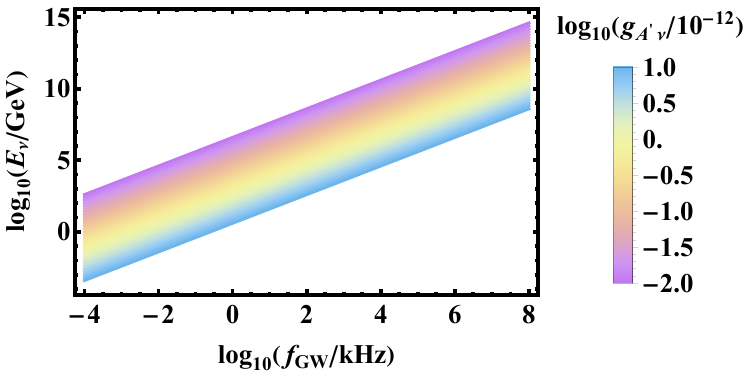}
\caption{The relation of gravitational wave frequency and neutrino flux energy with varying coupling (left) $g_{\phi\nu}$ and (right) $g_{A^{\prime}\nu}$. For both the cases we take $\alpha_g = 0.3$.}
\label{fvse}
\end{figure*}

The dependence of neutrino flux and energy along with GW strain and frequency on the source parameters, such as mass of the BH, gravitational fine structure constant, luminosity distance of the source, and spin of the BH are of pivotal importance as they can pave a way to connect the two messengers, GW and neutrino signals, which would otherwise seem to be independent of each other.
The left (right) panel of Fig.~\ref{fvse} is a bandplot of gravitational wave frequency and neutrino energy for different values of coupling for the scalar (vector) background case. For this we have considered the $\alpha_{g}=0.3$ and $a_*=0.9$. This plot shows that the gravitational wave frequency and neutrino energy have a linear dependence in the log scale. The energy-axis intercept of the graph corresponds to the coupling between the neutrinos and ultra-light boson ($g_{\phi\nu}$ or $g_{A^{\prime}\nu}$). It is also observable that the energy of neutrino and the coupling has an inverse relation, i.e. high energy neutrinos signify low coupling at constant gravitational wave frequency. This partially signifies how the two messengers can be put to work to understand the highly sought after coupling between the neutrinos and the ultra-light bosons.
\begin{figure*}
\centering
\includegraphics[scale=0.57]{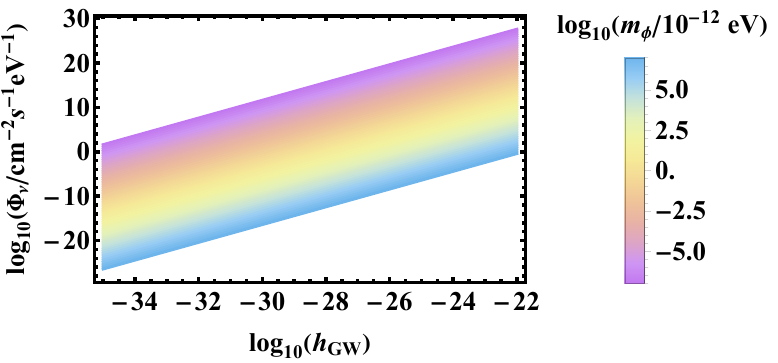}~~~
\includegraphics[scale=0.57]{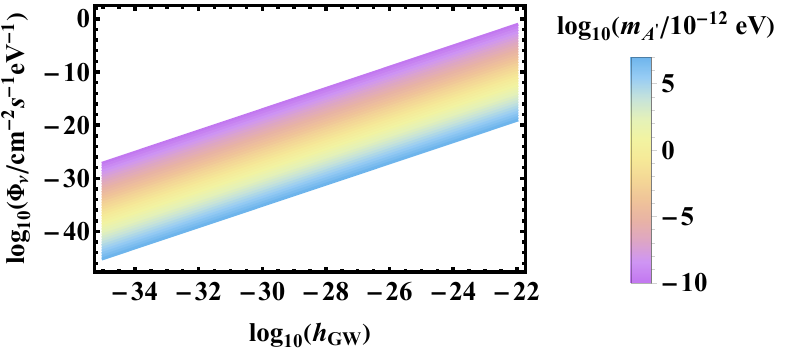}
\caption{The relation of gravitational wave strain and neutrino flux with varying mass of the ultra-light (left) scalar and (right) vector. It the left panel we take $g_{\phi\nu}=10^{-7}$ and for the right panel we take $g_{A^{\prime}\nu} = 10^{-10}$. For both the cases $\alpha_g = 0.3$.}
\label{hvsphi}
\end{figure*}
In the left (right) panel of Fig.~\ref{hvsphi}, we show the mutual dependence of the neutrino flux and the GW strain for the scalar (vector) background case. It is prominent that gravitational wave strain is directly proportional, in the logarithmic scale, to the neutrino flux generated and the intercept signifies the mass of the boson. Neutrino flux and mass of the ultralight boson have an inverse dependence, i.e. higher flux corresponds to lower mass for constant gravitational wave strain.

\subsection{Benchmark Cases}
\label{directsig}
\begin{figure*}
\centering
\includegraphics[scale=0.365]{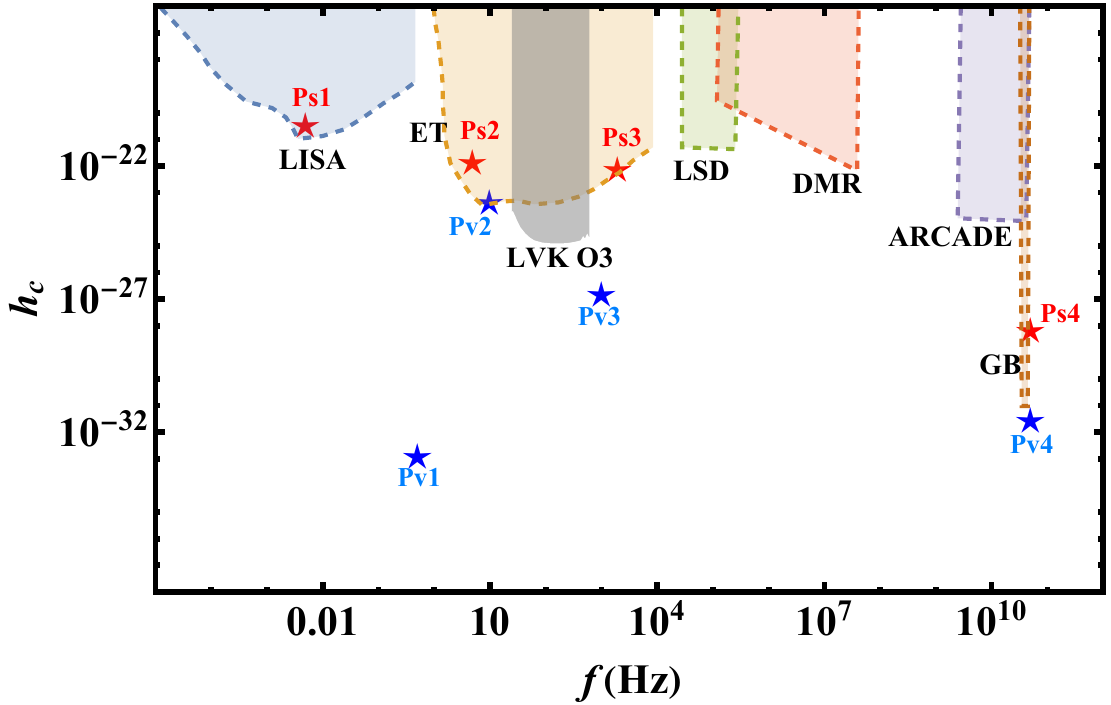}~~~~~~
\includegraphics[scale=0.4]{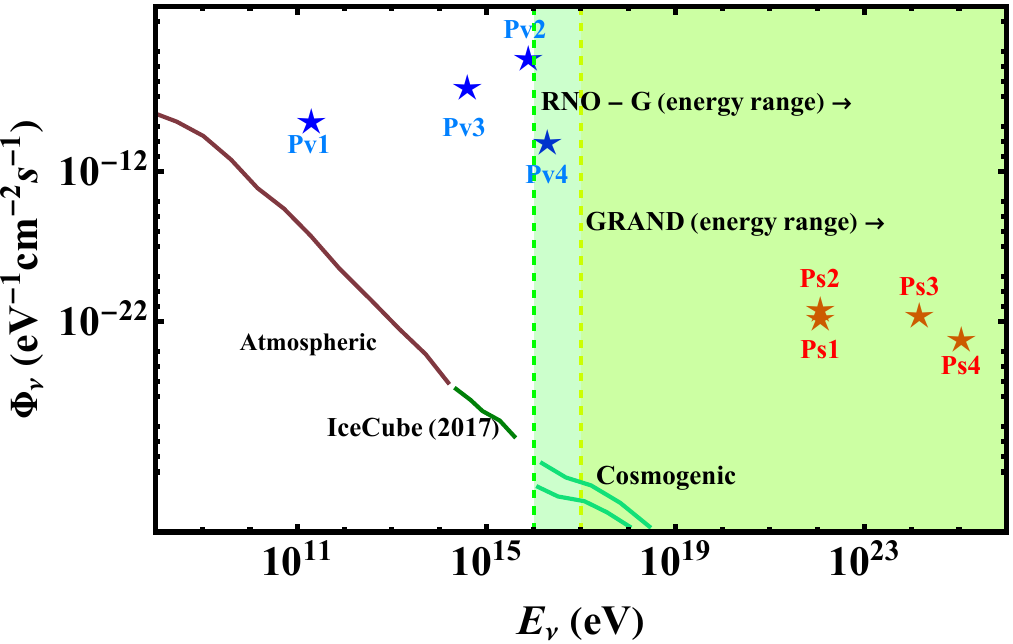}
\caption{\footnotesize (Left) We show the characteristic strain and the frequency of the GW signal arising from the balanced phase of the quenched superradiance mechanism from the benchmark points given in Tabs.~\ref{scalar__BP},~\ref{vector__BP}. The sensitivity curves of the future and proposed GW detectors have also been shown. The gray region denotes the bounds on the monochormatic GW from LIGO O3. (Right) We show the active neutrino flux and the neutrino energy arising from the balanced phase os the quenched superradiance mechanism from the benchmark points given in Tabs.~\ref{scalar__BP},~\ref{vector__BP}. For the purpose of comparison the diffuse neutrino flux from atmospheric neutrinos have been shown. Furthermore, the flux observed by ICECUBE and the predicted cosmogenic flux have also been shown. The yellow and green shaded regions show the detection strength of the experiment GRAND and RNO-G respectively at their completion. For the both the left and the right panel, the red and blue stars denote the scalar and vector benchmark parameters respectively.}
\label{sigbpres}
\end{figure*}

To further illustrate the versatility and the range of this method, we show a few specific scenarios for which both the signals will be relevant. We consider eight benchmark cases (four for the scalar case and four for the vector case) and discuss their imprint in the neutrino flux-energy and gravitational wave frequency-strain space. The benchmark cases for the scalar (vector) are shown in Tab. \ref{scalar__BP} (Tab. \ref{vector__BP}). For both scalar and vector case, the first benchmark point (BP) is motivated by supermassive BH, the second and third are motivated from astrophysical BH and the final one by PBH.
The distance of the BHs from the earth has been taken with a conservative approach, i.e., the shortest distance considered in the benchmark cases is much larger than the distance of the BH closest to earth which is at a distance of $480$ pc~\cite{El-Badry:2022zih}. Furthermore, the couplings that we consider in the BPs are consistent with the existing bounds~\cite{Kachelriess:2000qc,Farzan:2002wx,Forastieri:2019cuf} and the spin has considered aligned with the observed spin for Sgr $\mathrm{A}^*$~\cite{Daly:2023axh}. It is to be noted that the benchmark parameters are considered in such a way that the resulting GW and neutrino flux can be at a level which is in the reach of upcoming GW and neutrino detectors respectively.
\begin{table}[H]
\centering
\begin{tabular}{|c|c|c|c|c|}
\hline
BP ID & $M_{\mathrm{BH}}~(M_{\odot})$ & $g_{\phi\nu}$ & $a_*$ & $d~{(\mathrm{kpc})}$ \\ \hline \hline
   Ps1   &          $ 3.9\times 10^6  $      &   $ 10^{-9}$        &   $0.9$    &   $100$              \\ \hline
   Ps2   &             3900    &     $ 3.98\times 10^{-9}$        &   0.95    &    50            \\ \hline
   Ps3   &              9.8    &     $ 5\times 10^{-9} $      &   0.95    &   30               \\ \hline
   Ps4   &             $ 3.9\times 10^{-7}$   &    $  10^{-7}$        &   0.9    &   10              \\ \hline
\end{tabular}
\caption{The BPs for the scalar scenario.}
\label{scalar__BP}
\end{table}
\begin{table}[H]
\centering
\begin{tabular}{|c|c|c|c|c|}
\hline
BP ID & $M_{\mathrm{BH}}~(M_{\odot})$ & $g_{A^{\prime}\nu}$ & $a_*$ & $d~{(\mathrm{kpc})}$ \\ \hline \hline
   Pv1   &          $ 3.9\times 10^4  $      &   $ 10^{-12}$        &   $0.9$    &   $100$              \\ \hline
   Pv2   &             1950    &     $ 7.94\times 10^{-14}$        &   0.95    &    10            \\ \hline
   Pv3   &              19.5    &     $ 10^{-12} $      &   0.95    &   7               \\ \hline
   Pv4   &             $ 3.9\times 10^{-7}$   &    $  10^{-10}$        &   0.9    &   5              \\ \hline
\end{tabular}
\caption{The BPs for the vector scenario.}
\label{vector__BP}
\end{table}
The signals arising from the BPs given in Tabs.~\ref{scalar__BP} and \ref{vector__BP}, are shown in the left and right panels of Fig.~\ref{sigbpres} in the GW frequency-strain and neutrino flux-energy space, respectively. It is to be noted here that the scalar BPs are depicted with red `star' signs whereas the vector BPs are shown with blue star sign in Fig.~\ref{sigbpres}. It can be seen from Fig. ~\ref{sigbpres} that scalar BPs tend to lead to much higher energy neutrinos and high GW strain than vector BPs. This arises from the fundamental differences in the fermion production mechanism which leads to different production rates which in turn leads to different allowed values of critical field value (shown in Fig. \ref{condsca}) for the same masses of scalar and vectors. Now, we discuss the prospects of the detection of the signals arising from each of these BPs.
\begin{itemize}
\item \textbf{Ps1:} This leads to a gravitational wave signal with frequency $\mathcal{O}(0.01)$ Hz, which is in the range of LISA. The $\tau_{\mathrm{balanced}}$ for this case is approximately 330 years. Therefore, in calculating the characteristic GW strain for this BP, we take a LISA four years projection and the strain turns out to be $\mathcal{O}(10^{-21})$ which is within LISA sensitivity range. Furthermore, this BP also creates sterile neutrino flux with energy $\mathcal{O}(10^{22})$ eV and differential flux $\mathcal{O}(10^{-16})\mathrm{~eV^{-1}cm^{-2}s^{-1}}$, which in turn leads to a active neutrino flux of energy $\mathcal{O}(10^{22})$ eV and differential flux $\mathcal{O}(10^{-22})\mathrm{~eV^{-1}cm^{-2}s^{-1}}$. For all the scalar BPs, we consider the averaged probability of the transition from sterile to active neutrino over a large distance to be $\langle P_{sa}\rangle\sim 10^{-6}$, which is well within the constraints~\cite{Goldhagen:2021kxe}.
\item \textbf{Ps2:} This generates a GW signal with frequency $\mathcal{O}(10)$ Hz, which is in the range of ET. The $\tau_{\mathrm{balanced}}$ for this case is approximately 1.76 years for which the characteristic strain turns out to be $\mathcal{O}(10^{-22})$ which is within ET sensitivity range. Furthermore, this BP also creates active neutrino flux with energy $\mathcal{O}(10^{22})$ eV and differential flux $\mathcal{O}(10^{-22})\mathrm{~eV^{-1}cm^{-2}s^{-1}}$.
\item \textbf{Ps3:} This generates a GW signal with frequency $\mathcal{O}(1000)$ Hz, which is in the range of ET. The $\tau_{\mathrm{balanced}}$ for this case is approximately 8.53 seconds for which the characteristic strain turns out to be $\mathcal{O}(10^{-22})$ which is within ET sensitivity range. Furthermore, this BP also creates active neutrino flux with energy $\mathcal{O}(10^{24})$ eV and differential flux $\mathcal{O}(10^{-22})\mathrm{~eV^{-1}cm^{-2}s^{-1}}$.
\item \textbf{Ps4:} This leads to a GW signal with frequency $\mathcal{O}(10^{10})$ Hz. The $\tau_{\mathrm{balanced}}$ for this case is approximately 5.5 ns for which the characteristic strain turns out to be $\mathcal{O}(10^{-28})$ which is within the sensitivity range of the gaussian beam experiment. Furthermore, this BP also creates active neutrino flux with energy $\mathcal{O}(10^{25})$ eV and differential flux $\mathcal{O}(10^{-23})\mathrm{~eV^{-1}cm^{-2}s^{-1}}$.
The predicted  extremely high energy neutrinos from all the scalar BPs can in principle be visible through RNO-G or GRAND in the future.
\item \textbf{Pv1:} This leads to a GW strain with frequency $\mathcal{O}(1)$ Hz which is in the LISA range. Also, in this case, $\tau_{\mathrm{balanced}}$ is $\mathcal{O}(0.5)$ Tyr. Therefore, considering a four year projection of LISA, the characteristic strain of this signal is $\mathcal{O}(10^{-33})$ which is much lower than the capabilities of LISA, making this it infeasible for detection in the near future. However, the active neutrinos signal from this BP leads to a neutrino flux of $\mathcal{O}(10^{-9})\mathrm{~eV^{-1}cm^{-2}s^{-1}}$ with energy $\mathcal{O}(10^{11})$ eV which can be detected by super- Kamiokande.
\item \textbf{Pv2:} This generates a GW strain with frequency $\mathcal{O}(10)$ Hz which is in the ET range. Also, in this case, $\tau_{\mathrm{balanced}}$ is $\mathcal{O}(100)$ years. Therefore, considering a four year projection of ET, the characteristic strain of this signal is $\mathcal{O}(10^{-23})$ which is within the sensitivity of ET. Furthermore, this BP leads to a neutrino flux of $\mathcal{O}(10^{-5})\mathrm{~eV^{-1}cm^{-2}s^{-1}}$ with energy $\mathcal{O}(10^{16})$ eV which can be detected by IceCube and RNO-G.
\item \textbf{Pv3:} This generates a GW strain with frequency $\mathcal{O}(1000)$ Hz which is in the ET range. Also, in this case, $\tau_{\mathrm{balanced}}$ is $\mathcal{O}(1000)$ years. Therefore, considering a four year projection of ET, the characteristic strain of this signal is $\mathcal{O}(10^{-27})$ which is much beyond the sensitivity of ET. However, this BP leads to a neutrino flux of $\mathcal{O}(10^{-7})\mathrm{~eV^{-1}cm^{-2}s^{-1}}$ with energy $\mathcal{O}(10^{15})$ eV which can be detected by IceCube.
\item \textbf{Pv4:} This leads to a GW strain with frequency $\mathcal{O}(10^{10})$ Hz which is in the range of the gaussian beam experiment. However, in this case, $\tau_{\mathrm{balanced}}$ is $\mathcal{O}(0.1)$ seconds which gives us the characteristic strain of this signal as $\mathcal{O}(10^{-31})$ which is slightly beyond the sensitivity of the gaussian beam experiments. However, this BP leads to a neutrino flux of $\mathcal{O}(10^{-11})\mathrm{~eV^{-1}cm^{-2}s^{-1}}$ with energy $\mathcal{O}(10^{16})$ eV which can be detected by IceCube and RNO-G.
\end{itemize} 
It is to be noted that all these outcomes of the BPs have been depicted in the left and right panels of Fig.~\ref{sigbpres} and Tabs. \ref{scalarBPo} and \ref{vectorBPo}.

\begin{table}[H]
\centering
\begin{tabular}{|c|c|c|c|}
\hline
BP ID & $m_{\phi}$ (eV) & $h_{\mathrm{rms}}$ & $\tau_{\mathrm{balanced}}$ (s)  \\ \hline \hline
   Ps1   &          $ 10^{-17}  $      &   $5.37 \times 10^{-24}$        &   $5.5\times 10^{10}$               \\ \hline
   Ps2   &           $10^{-14}$    &     $ 1.074\times 10^{-26}$        &  $5.57\times 10^{7}$             \\ \hline
   Ps3   &           $3.98\times 10^{-12}$    &     $7.29\times 10^{-25} $      &   $8.53$                \\ \hline
   Ps4   &             $ 10^{-4}$   &   $5.37\times 10^{-30}$          &        $5.51\times 10^{-9}$       \\ \hline
\end{tabular}
\caption{The outcome of BPs for the scalar scenario.}
\label{scalarBPo}
\end{table}
\begin{table}[H]
\centering
\begin{tabular}{|c|c|c|c|}
\hline
BP ID & $m_{A^{\prime}}$ (eV) & $h_{\mathrm{rms}}$ & $\tau_{\mathrm{balanced}}$ (s)  \\ \hline \hline
   Pv1   &          $ 10^{-15}  $      &   $1.95 \times 10^{-37}$        &   $1.64\times 10^{20}$               \\ \hline
   Pv2   &           $2\times 10^{-14}$    &     $ 1.95\times 10^{-28}$        &  $5.513\times 10^{9}$             \\ \hline
   Pv3   &           $2\times 10^{-12}$    &     $5.57\times 10^{-33} $      &   $2.05\times 10^{10}$                \\ \hline
   Pv4   &             $ 10^{-4}$   &   $3.9\times 10^{-37}$          &        $0.164$       \\ \hline
\end{tabular}
\caption{The outcome of BPs for the vector scenario.}
\label{vectorBPo}
\end{table}
Through these figures and the tables, we show that future GW and high-energy neutrino detectors can lead to the simultaneous detection of these gravitational wave strain and neutrino flux which may open up new window to study ULBs.

\section{Summary and Conclusion}
\label{sec:concl}
In this work, we for the first time discuss a novel way of probing the quenched superradiance mechanism with simultaneous detection of GW and neutrino flux. The quenched superradiance of black holes is a mechanism where the superradiant energy extraction from a rotating black hole by the bosonic cloud is balanced by the energy extraction from the bosonic cloud by fermion provided that the fermions have the `right' coupling with the bosons. 
We consider the bosonic (scalar and vector) couplings of neutrinos where the boson (scalar  and vector) forms gravitational atom around the rotating black holes extracting its energy at the expense of its mass and spin. In this regard it is to be noted that due to the parametric production of neutrinos from the scalar cloud, one can not consider a direct coupling between the active neutrinos and the scalar as it leads to inefficient energy extraction from the scalar cloud rendering the quenched superradiance impossible. Hence, we consider a sterile neutrino $\nu_s$ to couple with the scalar, and after it is produced and propagates, a small part of the sterile neutrinos are converted into active neutrinos due to its small mixing with active neutrinos. On the other hand, for vector cloud the neutrinos are produced via Schwinger pair production, which is free of the above mentioned problem and hence we consider the lightest neutrino mass eigenstate $\nu_1$ to couple with the vector. 
After outlining the basics of the quenched superradiance mechanism, we show the region of parameter space in which the balance of energy extraction can occur. This is subject to two conditions on the field value of the bosons, i.e., the peak value should be much less than the peak value where the cloud mass is $10\%$ of the black hole mass, but it should be large enough to facilitate the production of the neutrinos. Next we discuss the time-lines involving the quenched superradiance mechanism which can be divided mainly into three parts, i.e., the growth phase where the boson cloud grows drawing energy from the black hole, the balanced phase where the boson cloud remains at a constant mass and the depletion phase where superradiant energy extraction stops yet fermion production continues and as a result the cloud depletes. We show our results as band-plots of the properties of the two signals to give an idea regarding what can be the nature of the simultaneous signals arising from this mechanism. Finally we consider a few benchmark points for both the scalar and vector cases and show the possible signals arising from them. We show that for our benchmarks, the neutrino signals coming from the scalars benchmarks will be in the ultra-high energy regime and can be probed in future with neutrino observatories like GRAND and RNO-G, whereas for the vector benchmarks the same can be probed via Super-Kamiokande and IceCube. Furthermore, we show that the gravitational waves arising from the scalar benchmarks are much higher and therefore much more detectable than the ones arising from the vector cases.
It is worth-mentioning that here we consider quenched superradiance in isolated black holes whereas the same can be considered for binary black hole systems as well, which may lead to novel features in the gravitational waves during their mergers. Furthermore, if the black holes that we consider are of primordial origin and can constitute some part of dark matter, then we can also predict a neutrino background from them provided that the black holes have sufficient spin and the neutrinos have right coupling with the bosons. We leave such possibilities for future work.
In conclusion, we believe that this study will open up the possibility of using quenched superradiance mechanism that model and predicts multi-messenger signals and could be proven to be of extreme importance in the upcoming times as the new generation of gravitational wave and neutrino detectors will be operative.

%
 

\vspace*{0.5cm}
\acknowledgments
IKB and SB thank Anna John for useful discussions. IKB acknowledges the support by the MHRD, Government of India, under the Prime Minister's Research Fellows (PMRF) Scheme, 2022. UKD acknowledges support from the Anusandhan National Research Foundation (ANRF), Government of India under Grant Reference No. CRG/2023/003769.

\bibliographystyle{h-physrev}
\bibliography{ULSMMA_ref.bib}

\begin{thebibliography}{100}

\bibitem{SAGE:1999zrn}
SAGE, D.~N. Abdurashitov {\em et~al.},
\newblock Nucl. Phys. B Proc. Suppl. {\bf 77}, 20 (1999).

\bibitem{MiniBooNE:2012meu}
MiniBooNE, SciBooNE, G.~Cheng {\em et~al.},
\newblock Phys. Rev. D {\bf 86}, 052009 (2012), 1208.0322.

\bibitem{SciBooNE:2011qyf}
SciBooNE, MiniBooNE, K.~B.~M. Mahn {\em et~al.},
\newblock Phys. Rev. D {\bf 85}, 032007 (2012), 1106.5685.

\bibitem{Borexino:2013zhu}
Borexino, G.~Bellini {\em et~al.},
\newblock Phys. Rev. D {\bf 89}, 112007 (2014), 1308.0443.

\bibitem{DUNE:2020fgq}
DUNE, B.~Abi {\em et~al.},
\newblock Eur. Phys. J. C {\bf 81}, 322 (2021), 2008.12769.

\bibitem{Hyper-Kamiokande:2018ofw}
Hyper-Kamiokande, K.~Abe {\em et~al.},
\newblock (2018), 1805.04163.

\bibitem{KamLAND:2014gul}
KamLAND, A.~Gando {\em et~al.},
\newblock Phys. Rev. C {\bf 92}, 055808 (2015), 1405.6190.

\bibitem{SNO:2011hxd}
SNO, B.~Aharmim {\em et~al.},
\newblock Phys. Rev. C {\bf 88}, 025501 (2013), 1109.0763.

\bibitem{MicroBooNE:2016pwy}
MicroBooNE, R.~Acciarri {\em et~al.},
\newblock JINST {\bf 12}, P02017 (2017), 1612.05824.

\bibitem{T2K:2011qtm}
T2K, K.~Abe {\em et~al.},
\newblock Nucl. Instrum. Meth. A {\bf 659}, 106 (2011), 1106.1238.

\bibitem{IceCube:2013low}
IceCube, M.~G. Aartsen {\em et~al.},
\newblock Science {\bf 342}, 1242856 (2013), 1311.5238.

\bibitem{IceCube:2013cdw}
IceCube, M.~G. Aartsen {\em et~al.},
\newblock Phys. Rev. Lett. {\bf 111}, 021103 (2013), 1304.5356.

\bibitem{ICAL:2015stm}
ICAL, S.~Ahmed {\em et~al.},
\newblock Pramana {\bf 88}, 79 (2017), 1505.07380.

\bibitem{RNO-G:2023kag}
RNO-G, J.~Henrichs {\em et~al.},
\newblock PoS {\bf ICRC2023}, 259 (2023).

\bibitem{NOvA:2007rmc}
NOvA, D.~S. Ayres {\em et~al.},
\newblock (2007).

\bibitem{Kamiokande:1986xwg}
Kamiokande, M.~Nakahata {\em et~al.},
\newblock J. Phys. Soc. Jap. {\bf 55}, 3786 (1986).

\bibitem{K2K:2000kji}
K2K, A.~Suzuki {\em et~al.},
\newblock Nucl. Instrum. Meth. A {\bf 453}, 165 (2000), hep-ex/0004024.

\bibitem{ANTARES:1999fhm}
ANTARES, E.~Aslanides {\em et~al.},
\newblock (1999), astro-ph/9907432.

\bibitem{MINOS:1998kez}
MINOS, I.~Ambats {\em et~al.},
\newblock (1998).

\bibitem{Anderson:1998zza}
K.~Anderson {\em et~al.},
\newblock (1998).

\bibitem{CHOOZ:1997cow}
CHOOZ, M.~Apollonio {\em et~al.},
\newblock Phys. Lett. B {\bf 420}, 397 (1998), hep-ex/9711002.

\bibitem{Super-Kamiokande:1998wen}
Super-Kamiokande, Y.~Fukuda {\em et~al.},
\newblock Phys. Lett. B {\bf 433}, 9 (1998), hep-ex/9803006.

\bibitem{IceCube-Gen2:2020qha}
IceCube-Gen2, M.~G. Aartsen {\em et~al.},
\newblock J. Phys. G {\bf 48}, 060501 (2021), 2008.04323.

\bibitem{DayaBay:2007fgu}
Daya Bay, X.~Guo {\em et~al.},
\newblock (2007), hep-ex/0701029.

\bibitem{Frejus:1994brq}
Frejus, K.~Daum {\em et~al.},
\newblock Z. Phys. C {\bf 66}, 417 (1995).

\bibitem{RENO:2018dro}
RENO, G.~Bak {\em et~al.},
\newblock Phys. Rev. Lett. {\bf 121}, 201801 (2018), 1806.00248.

\bibitem{Andres:1999hm}
E.~Andres {\em et~al.},
\newblock Astropart. Phys. {\bf 13}, 1 (2000), astro-ph/9906203.

\bibitem{DeBonis:2014jlo}
I.~De~Bonis {\em et~al.},
\newblock (2014), 1409.4405.

\bibitem{KM3NeT:2018wnd}
KM3NeT, S.~Aiello {\em et~al.},
\newblock Astropart. Phys. {\bf 111}, 100 (2019), 1810.08499.

\bibitem{GRAND:2018iaj}
GRAND, J.~\'Alvarez-Mu\~niz {\em et~al.},
\newblock Sci. China Phys. Mech. Astron. {\bf 63}, 219501 (2020), 1810.09994.

\bibitem{Waxman:1997ti}
E.~Waxman and J.~N. Bahcall,
\newblock Phys. Rev. Lett. {\bf 78}, 2292 (1997), astro-ph/9701231.

\bibitem{Waxman:1998yy}
E.~Waxman and J.~N. Bahcall,
\newblock Phys. Rev. D {\bf 59}, 023002 (1999), hep-ph/9807282.

\bibitem{Super-Kamiokande:2002hei}
Super-Kamiokande, M.~Malek {\em et~al.},
\newblock Phys. Rev. Lett. {\bf 90}, 061101 (2003), hep-ex/0209028.

\bibitem{Schramm:1980xv}
D.~N. Schramm and G.~Steigman,
\newblock Astrophys. J. {\bf 243}, 1 (1981).

\bibitem{Barr:1989ru}
G.~Barr, T.~K. Gaisser, and T.~Stanev,
\newblock Phys. Rev. D {\bf 39}, 3532 (1989).

\bibitem{Bellini:2013wsa}
G.~Bellini, A.~Ianni, L.~Ludhova, F.~Mantovani, and W.~F. McDonough,
\newblock Prog. Part. Nucl. Phys. {\bf 73}, 1 (2013), 1310.3732.

\bibitem{Mirizzi:2015eza}
A.~Mirizzi {\em et~al.},
\newblock Riv. Nuovo Cim. {\bf 39}, 1 (2016), 1508.00785.

\bibitem{Davis:1968cp}
R.~Davis, Jr., D.~S. Harmer, and K.~C. Hoffman,
\newblock Phys. Rev. Lett. {\bf 20}, 1205 (1968).

\bibitem{Super-Kamiokande:2016yck}
Super-Kamiokande, K.~Abe {\em et~al.},
\newblock Phys. Rev. D {\bf 94}, 052010 (2016), 1606.07538.

\bibitem{SNO:2002tuh}
SNO, Q.~R. Ahmad {\em et~al.},
\newblock Phys. Rev. Lett. {\bf 89}, 011301 (2002), nucl-ex/0204008.

\bibitem{Ciscar-Monsalvatje:2024tvm}
M.~C\'\i{}scar-Monsalvatje, G.~Herrera, and I.~M. Shoemaker,
\newblock (2024), 2402.00985.

\bibitem{Herrera:2024upj}
G.~Herrera, S.~Horiuchi, and X.~Qi,
\newblock (2024), 2405.14946.

\bibitem{Griest:1989wd}
K.~Griest and M.~Kamionkowski,
\newblock Phys. Rev. Lett. {\bf 64}, 615 (1990).

\bibitem{Ferreira:2020fam}
E.~G.~M. Ferreira,
\newblock Astron. Astrophys. Rev. {\bf 29}, 7 (2021), 2005.03254.

\bibitem{Arvanitaki:2009fg}
A.~Arvanitaki, S.~Dimopoulos, S.~Dubovsky, N.~Kaloper, and J.~March-Russell,
\newblock Phys. Rev. D {\bf 81}, 123530 (2010), 0905.4720.

\bibitem{Peebles:2002gy}
P.~J.~E. Peebles and B.~Ratra,
\newblock Rev. Mod. Phys. {\bf 75}, 559 (2003), astro-ph/0207347.

\bibitem{Fischbach:1992nm}
E.~Fischbach, G.~T. Gillies, D.~E. Krause, J.~G. Schwan, and C.~Talmadge,
\newblock Metrologia {\bf 29}, 213 (1992).

\bibitem{Carroll:1989vb}
S.~M. Carroll, G.~B. Field, and R.~Jackiw,
\newblock Phys. Rev. D {\bf 41}, 1231 (1990).

\bibitem{Carroll:1991zs}
S.~M. Carroll and G.~B. Field,
\newblock Phys. Rev. D {\bf 43}, 3789 (1991).

\bibitem{Harari:1992ea}
D.~Harari and P.~Sikivie,
\newblock Phys. Lett. B {\bf 289}, 67 (1992).

\bibitem{Raffelt:1987im}
G.~Raffelt and L.~Stodolsky,
\newblock Phys. Rev. D {\bf 37}, 1237 (1988).

\bibitem{Arvanitaki:2010sy}
A.~Arvanitaki and S.~Dubovsky,
\newblock Phys. Rev. D {\bf 83}, 044026 (2011), 1004.3558.

\bibitem{Chen:2023vkq}
Y.~Chen, X.~Xue, and V.~Cardoso,
\newblock JCAP {\bf 02}, 035 (2025), 2308.00741.

\bibitem{Gelmini:1980re}
G.~B. Gelmini and M.~Roncadelli,
\newblock Phys. Lett. B {\bf 99}, 411 (1981).

\bibitem{Georgi:1974sy}
H.~Georgi and S.~L. Glashow,
\newblock Phys. Rev. Lett. {\bf 32}, 438 (1974).

\bibitem{Pati:1974yy}
J.~C. Pati and A.~Salam,
\newblock Phys. Rev. D {\bf 10}, 275 (1974),
\newblock [Erratum: Phys.Rev.D 11, 703--703 (1975)].

\bibitem{Mohapatra:1974hk}
R.~N. Mohapatra and J.~C. Pati,
\newblock Phys. Rev. D {\bf 11}, 566 (1975).

\bibitem{Goodsell:2009xc}
M.~Goodsell, J.~Jaeckel, J.~Redondo, and A.~Ringwald,
\newblock JHEP {\bf 11}, 027 (2009), 0909.0515.

\bibitem{Fabbrichesi:2020wbt}
M.~Fabbrichesi, E.~Gabrielli, and G.~Lanfranchi,
\newblock (2020), 2005.01515.

\bibitem{Foot:1990mn}
R.~Foot,
\newblock Mod. Phys. Lett. A {\bf 06}, 527 (1991).

\bibitem{He:1991qd}
X.-G. He, G.~C. Joshi, H.~Lew, and R.~R. Volkas,
\newblock Phys. Rev. D {\bf 44}, 2118 (1991).

\bibitem{Foot:1994vd}
R.~Foot, X.~G. He, H.~Lew, and R.~R. Volkas,
\newblock Phys. Rev. D {\bf 50}, 4571 (1994), hep-ph/9401250.

\bibitem{XENON100:2014csq}
XENON100, E.~Aprile {\em et~al.},
\newblock Phys. Rev. D {\bf 90}, 062009 (2014), 1404.1455,
\newblock [Erratum: Phys.Rev.D 95, 029904 (2017)].

\bibitem{XENON:2019gfn}
XENON, E.~Aprile {\em et~al.},
\newblock Phys. Rev. Lett. {\bf 123}, 251801 (2019), 1907.11485.

\bibitem{XENON:2020rca}
XENON, E.~Aprile {\em et~al.},
\newblock Phys. Rev. D {\bf 102}, 072004 (2020), 2006.09721.

\bibitem{Ehret:2010mh}
K.~Ehret {\em et~al.},
\newblock Phys. Lett. B {\bf 689}, 149 (2010), 1004.1313.

\bibitem{Garcia:2020qrp}
A.~A. Garcia, K.~Bondarenko, S.~Ploeckinger, J.~Pradler, and A.~Sokolenko,
\newblock JCAP {\bf 10}, 011 (2020), 2003.10465.

\bibitem{Witte:2020rvb}
S.~J. Witte, S.~Rosauro-Alcaraz, S.~D. McDermott, and V.~Poulin,
\newblock JHEP {\bf 06}, 132 (2020), 2003.13698.

\bibitem{Redondo:2013lna}
J.~Redondo and G.~Raffelt,
\newblock JCAP {\bf 08}, 034 (2013), 1305.2920.

\bibitem{An:2013yfc}
H.~An, M.~Pospelov, and J.~Pradler,
\newblock Phys. Lett. B {\bf 725}, 190 (2013), 1302.3884.

\bibitem{Joshipura:2003jh}
A.~S. Joshipura and S.~Mohanty,
\newblock Phys. Lett. B {\bf 584}, 103 (2004), hep-ph/0310210.

\bibitem{KumarPoddar:2020kdz}
T.~Kumar~Poddar, S.~Mohanty, and S.~Jana,
\newblock Eur. Phys. J. C {\bf 81}, 286 (2021), 2002.02935.

\bibitem{KumarPoddar:2019ceq}
T.~Kumar~Poddar, S.~Mohanty, and S.~Jana,
\newblock Phys. Rev. D {\bf 100}, 123023 (2019), 1908.09732.

\bibitem{Bekenstein:1973mi}
J.~D. Bekenstein,
\newblock Phys. Rev. D {\bf 7}, 949 (1973).

\bibitem{Bekenstein:1998nt}
J.~D. Bekenstein and M.~Schiffer,
\newblock Phys. Rev. D {\bf 58}, 064014 (1998), gr-qc/9803033.

\bibitem{Detweiler:1980uk}
S.~L. Detweiler,
\newblock Phys. Rev. D {\bf 22}, 2323 (1980).

\bibitem{LIGOScientific:2016aoc}
LIGO Scientific, Virgo, B.~P. Abbott {\em et~al.},
\newblock Phys. Rev. Lett. {\bf 116}, 061102 (2016), 1602.03837.

\bibitem{NANOGrav:2023gor}
NANOGrav, G.~Agazie {\em et~al.},
\newblock Astrophys. J. Lett. {\bf 951}, L8 (2023), 2306.16213.

\bibitem{NANOGrav:2023hde}
NANOGrav, G.~Agazie {\em et~al.},
\newblock Astrophys. J. Lett. {\bf 951}, L9 (2023), 2306.16217.

\bibitem{EPTA:2023fyk}
EPTA, InPTA:, J.~Antoniadis {\em et~al.},
\newblock Astron. Astrophys. {\bf 678}, A50 (2023), 2306.16214.

\bibitem{EPTA:2023sfo}
EPTA, J.~Antoniadis {\em et~al.},
\newblock Astron. Astrophys. {\bf 678}, A48 (2023), 2306.16224.

\bibitem{EPTA:2023xxk}
EPTA, J.~Antoniadis {\em et~al.},
\newblock (2023), 2306.16227.

\bibitem{Reardon:2023gzh}
D.~J. Reardon {\em et~al.},
\newblock Astrophys. J. Lett. {\bf 951}, L6 (2023), 2306.16215.

\bibitem{Zic:2023gta}
A.~Zic {\em et~al.},
\newblock Publ. Astron. Soc. Austral. {\bf 40}, e049 (2023), 2306.16230.

\bibitem{Reardon:2023zen}
D.~J. Reardon {\em et~al.},
\newblock Astrophys. J. Lett. {\bf 951}, L7 (2023), 2306.16229.

\bibitem{Xu:2023wog}
H.~Xu {\em et~al.},
\newblock Res. Astron. Astrophys. {\bf 23}, 075024 (2023), 2306.16216.

\bibitem{LISA:2017pwj}
LISA, P.~Amaro-Seoane {\em et~al.},
\newblock (2017), 1702.00786.

\bibitem{Ruan:2018tsw}
W.-H. Ruan, Z.-K. Guo, R.-G. Cai, and Y.-Z. Zhang,
\newblock Int. J. Mod. Phys. A {\bf 35}, 2050075 (2020), 1807.09495.

\bibitem{Kawamura:2011zz}
S.~Kawamura {\em et~al.},
\newblock Class. Quant. Grav. {\bf 28}, 094011 (2011).

\bibitem{Phinney:2004bbo}
S.~P. et~al.,
\newblock NASA Mission Concept Study  (2004).

\bibitem{Punturo:2010zz}
M.~Punturo {\em et~al.},
\newblock Class. Quant. Grav. {\bf 27}, 194002 (2010).

\bibitem{Reitze:2019iox}
D.~Reitze {\em et~al.},
\newblock Bull. Am. Astron. Soc. {\bf 51}, 035 (2019), 1907.04833.

\bibitem{Arvanitaki:2012cn}
A.~Arvanitaki and A.~A. Geraci,
\newblock Phys. Rev. Lett. {\bf 110}, 071105 (2013), 1207.5320.

\bibitem{Aggarwal:2020umq}
N.~Aggarwal {\em et~al.},
\newblock Phys. Rev. Lett. {\bf 128}, 111101 (2022), 2010.13157.

\bibitem{Gertsenshtein:1962gw}
M.~E. Gertsenshtein,
\newblock Soviet Physics JETP {\bf 14}, 84 (1962).

\bibitem{Bahre:2013ywa}
R.~B\"ahre {\em et~al.},
\newblock JINST {\bf 8}, T09001 (2013), 1302.5647.

\bibitem{Albrecht:2020ntd}
C.~Albrecht {\em et~al.},
\newblock EPJ Tech. Instrum. {\bf 8}, 5 (2021), 2004.13441.

\bibitem{Chaudhuri:2014dla}
S.~Chaudhuri {\em et~al.},
\newblock Phys. Rev. D {\bf 92}, 075012 (2015), 1411.7382.

\bibitem{Silva-Feaver:2016qhh}
M.~Silva-Feaver {\em et~al.},
\newblock IEEE Trans. Appl. Supercond. {\bf 27}, 1400204 (2017), 1610.09344.

\bibitem{Holometer:2016qoh}
Holometer, A.~S. Chou {\em et~al.},
\newblock Phys. Rev. D {\bf 95}, 063002 (2017), 1611.05560.

\bibitem{Akutsu:2008qv}
T.~Akutsu {\em et~al.},
\newblock Phys. Rev. Lett. {\bf 101}, 101101 (2008), 0803.4094.

\bibitem{Li:2003tv}
F.-Y. Li, M.-X. Tang, and D.-P. Shi,
\newblock Phys. Rev. D {\bf 67}, 104008 (2003), gr-qc/0306092.

\bibitem{Li:2004df}
F.-Y. Li and N.~Yang,
\newblock Chin. Phys. Lett. {\bf 21}, 2113 (2004), gr-qc/0410060.

\bibitem{Li:2006sx}
F.~Li, R.~M.~L. Baker, Jr., and Z.~Chen,
\newblock (2006), gr-qc/0604109.

\bibitem{Fixsen:2009xn}
D.~J. Fixsen {\em et~al.},
\newblock Astrophys. J. {\bf 734}, 5 (2011), 0901.0555.

\bibitem{Bowman:2018yin}
J.~D. Bowman, A.~E.~E. Rogers, R.~A. Monsalve, T.~J. Mozdzen, and N.~Mahesh,
\newblock Nature {\bf 555}, 67 (2018), 1810.05912.

\bibitem{Brito:2015oca}
R.~Brito, V.~Cardoso, and P.~Pani,
\newblock Lect. Notes Phys. {\bf 906}, pp.1 (2015), 1501.06570.

\bibitem{Zhang:2019eid}
J.~Zhang and H.~Yang,
\newblock Phys. Rev. D {\bf 101}, 043020 (2020), 1907.13582.

\bibitem{Cao:2023fyv}
Y.~Cao and Y.~Tang,
\newblock Phys. Rev. D {\bf 108}, 123017 (2023), 2307.05181.

\bibitem{Tomaselli:2024dbw}
G.~M. Tomaselli, T.~F.~M. Spieksma, and G.~Bertone,
\newblock Phys. Rev. Lett. {\bf 133}, 121402 (2024), 2407.12908.

\bibitem{Yoshino:2013ofa}
H.~Yoshino and H.~Kodama,
\newblock PTEP {\bf 2014}, 043E02 (2014), 1312.2326.

\bibitem{Baryakhtar:2017ngi}
M.~Baryakhtar, R.~Lasenby, and M.~Teo,
\newblock Phys. Rev. D {\bf 96}, 035019 (2017), 1704.05081.

\bibitem{El-Badry:2022zih}
K.~El-Badry {\em et~al.},
\newblock Mon. Not. Roy. Astron. Soc. {\bf 518}, 1057 (2023), 2209.06833.

\bibitem{Kachelriess:2000qc}
M.~Kachelriess, R.~Tomas, and J.~W.~F. Valle,
\newblock Phys. Rev. D {\bf 62}, 023004 (2000), hep-ph/0001039.

\bibitem{Farzan:2002wx}
Y.~Farzan,
\newblock Phys. Rev. D {\bf 67}, 073015 (2003), hep-ph/0211375.

\bibitem{Forastieri:2019cuf}
F.~Forastieri, M.~Lattanzi, and P.~Natoli,
\newblock Phys. Rev. D {\bf 100}, 103526 (2019), 1904.07810.

\bibitem{Daly:2023axh}
R.~A. Daly {\em et~al.},
\newblock Mon. Not. Roy. Astron. Soc. {\bf 527}, 428 (2023), 2310.12108.

\bibitem{Goldhagen:2021kxe}
K.~Goldhagen, M.~Maltoni, S.~E. Reichard, and T.~Schwetz,
\newblock Eur. Phys. J. C {\bf 82}, 116 (2022), 2109.14898.

\end{thebibliography}

\end{document}